# Systematic application of the M3Y *NN* forces for describing the capture process in heavy-ion collisions involving deformed target nuclei

I. I. Gontchar[1], M. V. Chushnyakova[2] and O. M. Sukhareva[2,*]

[1]*Physics and Chemistry Department, Omsk State Transport University, 644046 Omsk, Russia*

[2]*Physics Department, Omsk State Technical University, 644050 Omsk, Russia*

*e-mail:* o.m.sukhareva@gmail.com

We present results of a systematic study of the capture process through the barrier penetration model. The nucleus-nucleus interaction potential is calculated using the double-folding model (DFM) with the M3Y Paris NN forces. The nucleon densities entering the model are generated from the experimental three-parameter Fermi charge densities. The DFM has been extended to the case of deformed target nuclei. It is shown that the density-dependent M3Y NN forces with the finite range exchange part can be mimicked successfully by the zero-range density-independent forces. The latter option significantly reduces the required computer time. For the nucleon densities and target nuclei deformations we employ the values from the commonly used data bases. Thus, we do not vary any parameters to reach a better agreement with the data. The resulting cross-sections are compared with data for 20 reactions with the product of the charge numbers $Z_P \cdot Z_T$ ranging from 216 up to 2576. We discuss the opinion met in the literature that the M3Y NN forces provide a poorer description of the capture cross-sections in heavy-ion collisions in comparison to the NN forces coming from the relativistic mean-field approach. Our calculations show that the M3Y NN forces give an agreement with the data which is not perfect yet is not worse than the one resulting from the RMF NN forces.

## I. INTRODUCTION

The optical model is known to be one of the best tools for describing collisions of heavy ions including the coupled channels. The nucleus-nucleus potential entering the optical model is often based on the effective *NN* forces [1]. For this aim, the single folding approach was first developed for the case of proton impinging on a nucleus [2]. Later, this approach was generalized for the case of two complex colliding nuclei becoming the Double Folding Model (DFM) [3,4]. Since then, the DFM has been used widely for finding the real part of the optical potential and the Coulomb barriers (see, e.g. [5–8]).

For the *NN* forces, in the non-relativistic approximation, three versions are typically used. These are the zero-range Migdal forces [3,9] and two kinds of the Yukawa-type M3Y-forces: the Reid forces [6,10] and the Paris forces [11]. For calculating the capture cross-sections, the use of the Reid forces in comparison with the Paris forces was studied in [12], whereas the comparison between the Migdal- and M3Y Paris forces in the same context was performed in [13,14].

Recently, the Yukawa-type *NN* forces coming from the relativistic mean-field approach (RMF *NN* forces) were applied for calculating the fusion cross-sections of heavy ions [15–17]. In Refs. [15,16], it was claimed that the RMF *NN* forces provide "... a better choice than the M3Y interaction for fusion reaction considered in the entire range of barrier energies in predicting the cross-sections...". The purpose of the present work is to confirm or disprove this statement. For this goal, using the DFM we calculated the capture cross-sections for 20 reactions involving the deformed target nuclei. This model is described in Section II for the M3Y and RMF *NN* forces. The logic of our calculations with the M3Y *NN* forces is presented in Section III. In Section IV, the choice of reactions is explained. The nucleon densities entering the DFM are discussed in Section V. In Section VI, the cross-sections resulting from the M3Y Paris *NN* forces are compared with the experimental data and with the cross-sections of Ref. [15] resulting from the RMF *NN* forces. The conclusions are formulated in Section VII. Appendix contains the details of the DFM calculations for the case of the deformed target nucleus.

## II. THE DOUBLE FOLDING MODEL

Within the framework of the DFM, the Strong nucleus-nucleus Potential (SnnP) $U_n(\vec{R})$ comprises two parts, the direct $U_{nD}(\vec{R})$ and exchange $U_{nE}(\vec{R})$ ones:

$$U_n(\vec{R}) = U_{nD}(\vec{R}) + U_{nE}(\vec{R}). \qquad (1)$$

The vector $\vec{R}$ connects the centers of mass of the colliding nuclei. The direct part is evaluated as follows

$$U_{nD}(\vec{R}) = \int d\vec{r}_P \int d\vec{r}_T \rho_{AP}(\vec{r}_P) F(\rho_{FA}) v_D(s) \rho_{AT}(\vec{r}_T). \qquad (2)$$

Here $\vec{r}_P$ and $\vec{r}_T$ are the radius-vectors of the interacting points of the projectile (P) and target (T) nuclei. The distance between these points is defined by the vector $\vec{s} = \vec{r}_T - \vec{R} - \vec{r}_P$. The geometry of the collision and all these vectors are shown in Fig. 1. In Eq. (2), $\rho_{AP}$ and $\rho_{AT}$ represent the nucleon densities (diagonal components of the density matrixes) for the projectile and target nuclei, respectively. The multiplier $F(\rho_{FA})$ represents the density dependence of the NN forces and is the same for both direct and exchange parts. For the case of two spherical colliding nuclei, it reads [18]

$$F(\rho_{FA}) = 0.3429\{1 + 3.0232 \exp(-\beta\rho_{FA}) - \gamma\rho_{FA}\}, \tag{3}$$

$\beta = 3.5512$ fm$^3$, $\gamma = 0.5$ fm$^3$. This parameter set results in the lowest barrier (see [12] for details). The nucleon density $\rho_{FA}$ in Eq. (2) reads [19]

$$\rho_{FA} = \rho_{AP}(r_P) + \rho_{AT}(r_T). \tag{4}$$

This choice is dictated more by numerical convenience than by physical arguments.

The direct part of the NN forces comprises two Yukawa terms:

$$v_D(s) = \sum_{i=1}^{2} G_{Di} \frac{\exp(-s/r_{vi})}{s/r_{vi}}. \tag{5}$$

Parameters of the NN forces are presented in Table I. In the present work, we discuss only M3Y Paris forces calling them the NN forces.

Table I. Parameter set for the effective M3Y Paris NN forces. The coefficients $G_{Di}$, $G_{Efi}$ were obtained by fitting the G-matrix elements at three selected distances $r_{vi}$.

| Parameter | Value |
|---|---|
| $G_{D1}$ (MeV) | 11062 |
| $G_{D2}$ (MeV) | -2537.5 |
| $G_{Ef1}$ (MeV) | -1524.25 |
| $G_{Ef2}$ (MeV) | -518.75 |
| $G_{Ef3}$ (MeV) | -7.847 |
| $G_{E\delta 0}$ (MeV·fm$^3$) | -592 |
| $r_{v1}$ (fm) | 0.25 |
| $r_{v2}$ (fm) | 0.40 |
| $r_{v3}$ (fm) | 1.414 |

The exchange part of the SnnP reads

$$U_{nE}(\vec{R}) = \int d\vec{r}_P \int d\vec{r}_T \hat{\rho}_{AP}(\vec{r}_P; \vec{r}_P + \vec{s}) F(\rho_{FA}) v_E(s) \hat{\rho}_{AT}(\vec{r}_T; \vec{r}_T - \vec{s}) \exp(i\vec{k}_{rel}\vec{s}\, m_n/m_R). \tag{6}$$

Here $m_R$ is the reduced mass; $\hat{\rho}_{AP}$ ($\hat{\rho}_{AT}$) is the non-diagonal matrix element for the projectile (target) nucleus. These elements are evaluated using the density matrix expansion method of Refs. [20,21]:

$$\hat{\rho}_{AT}(\vec{r}_T; \vec{r}_T - \vec{s}) \approx \rho_{AT}(\vec{r}_T - \vec{s}/2) \tilde{\jmath}_1[k_{eff}(\vec{r}_T - \vec{s}/2) \cdot s], \tag{7}$$

$$\hat{\rho}_{AP}(\vec{r}_P; \vec{r}_P + \vec{s}) \approx \rho_{AP}(\vec{r}_P + \vec{s}/2) \tilde{\jmath}_1[k_{eff}(\vec{r}_P + \vec{s}/2) \cdot s], \tag{8}$$

$$\tilde{\jmath}_1(x) = 3[\sin(x) - x\cos(x)]/x^3. \tag{9}$$

For the effective Fermi momentum $k_{eff}$, we apply the simplest Slater approximation

$$k_{eff}(\vec{r}) = \left[\frac{3\pi^2 \rho_A(\vec{r})}{2}\right]^{2/3}. \tag{10}$$

The nucleon density $\rho_{FA}$ in Eq. (6) reads [6,18,22]

$$\rho_{FA} = \rho_{AP}\left(\left|\vec{r}_P + \frac{\vec{s}}{2}\right|\right) + \rho_{AT}\left(\left|\vec{r}_T - \frac{\vec{s}}{2}\right|\right). \tag{11}$$

This corresponds to the middle point between the interacting points of the projectile and target nuclei and is physically justified (see Fig. 1).

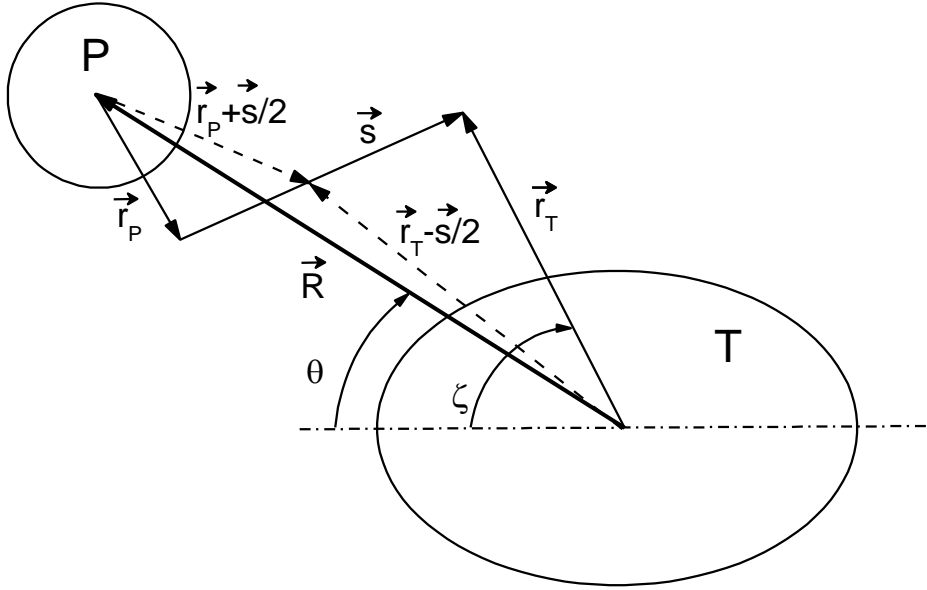

FIG. 1. Coordinate system used in the double-folding model.

The wave number corresponding to the relative motion of the colliding nuclei, $k_{rel}$, reads

$$k_{rel}(\vec{R}) = \sqrt{2m_R \left[E_{c.m.} - U_n(\vec{R}) - U_C(\vec{R})\right]}/\hbar. \tag{12}$$

Here $E_{c.m.}$ is the collision energy and $U_C(\vec{R})$ is the Coulomb interaction energy. The total effective interaction energy of the colliding nuclei $U(\vec{R})$ including in addition the centrifugal (rotational) energy $U_{rot}(\vec{R})$ reads:

$$U(\vec{R}) = U_n(\vec{R}) + U_C(\vec{R}) + U_{rot}(\vec{R}). \tag{13}$$

The exchange part of the *NN* forces $v_E(s)$ is met in the literature in two versions: with zero range and with finite (non-zero) range. Let us denote the former one as $\delta NN$ forces and the latter one as *fNN* forces. The simpler $\delta NN$ forces read:

$$v_{E\delta}(s) = G_{E\delta}\,\delta(\vec{s}). \tag{14}$$

In this case, the non-diagonal matrix elements in Eq. (6) are reduced to the diagonal elements and the integrals take the form of Eq. (2). Such integrals are evaluated rather fast by means of the Fourier transform (see details in [22–24]). Since in the present work we will modify $G_{E\delta}$, we denote the original value corresponding to [11] as $G_{E\delta 0}$.

The much more sophisticated *fNN* forces involve three Yukawa terms:

$$v_{Ef}(s) = \sum_{i=1}^{3} G_{Efi}\,\frac{\exp(-s/r_{vi})}{s/r_{vi}}. \tag{15}$$

In this case, evaluating the integrals in Eq. (6) is significantly more computer time-consuming. The required computer time dramatically increases even more when one deals with the deformed target nucleus because one needs to evaluate the integrals (6) many times for many values of $R$ and $\theta$ (see Fig. 1). On the other hand, only the *fNN* density-dependent forces (*fDDNN* forces) provide the correct saturated nucleon density for cold nuclear matter [18]. Moreover, the *fDDNN* forces have been applied only to the case of spherical colliding nuclei [4,12,19,25–29] whereas, in the present work, we aim to study the reactions with deformed target nuclei.

In addition, we would like to compare our results with the calculations obtained in [15] within the DFM with the relativistic mean field (RMF) *NN* forces. It has the same structure as in the case of M3Y *NN* forces. Most of the differences come from the way the *NN* forces are obtained. Starting from the standard Lagrangian density [15,30], the effective RMF *NN* interaction is constructed in the form of the Yukawa-type terms [15,17,31–33]

$$v_{RMF}(r) = \frac{g_\omega^2}{4\pi}\frac{\exp(-m_\omega rc/\hbar)}{r} + \frac{g_\rho^2}{4\pi}\frac{\exp(-m_\rho rc/\hbar)}{r} - \frac{g_\sigma^2}{4\pi}\frac{\exp(-m_\sigma rc/\hbar)}{r}$$
$$+ \frac{g_2^2}{4\pi}r\exp(-2m_\sigma rc/\hbar) + \frac{g_3^2}{4\pi}\frac{\exp(-3m_\sigma rc/\hbar)}{r} - J_{00}\delta(\vec{r}). \quad (16)$$

The free parameters of this theory, the meson masses $m_\rho$, $m_\sigma$, $m_\omega$ and the interaction strengths $g_\rho$, $g_\sigma$, $g_\omega$, $g_2$, $g_3$ are fitted to reproduce the static properties of the nuclei.

Thus, though formally similar, the M3Y and RMF *NN* forces come from different considerations. Moreover, the strength of the zero-range potential in the M3Y forces (Eqs. (14), (12A)) was scaled to reproduce the exchange integral of the potential obtained with the finite-range exchange term of the M3Y interaction. This has nothing to do with the effective interaction derived from RMF. Therefore, adding the last delta-function term in Eq. (16) as it was done in [15,16,33] seems to be questionable.

Moreover, the coupling strengths of different mesons in the RMF were carefully chosen to describe the saturation of nuclear matter as well as the ground-state structure of some nuclei. Therefore, the single nucleon exchange effects are already accounted for implicitly. It looks like adding by hand a zero-range potential to the in-medium *NN* interaction derived from the RMF meson fields is a sort of double-counting. Note that in [31,32] the zero-range pseudopotential absents.

These effective forces are folded with the densities analogously to Eq. (2). It should be noted that the RMF *NN* forces were applied in the literature only for the case of spherical target and projectile nuclei. Therefore, the SnnP reads

$$U_n(R) = \int d\vec{r}_P \int d\vec{r}_T \rho_{AP}(r_P) v_{RMF}(s) \rho_{AT}(r_T). \quad (17)$$

## III. COMPUTATIONAL ROUTINE

In our investigation, we use two computer codes: DFMSPH [24,34] and DFMDEF [35,36] (see details of the DFMDEF code in Appendix). The first one is designed for calculating the interaction energy of two spherical colliding nuclei and allows accounting for the density dependence of *NN* forces. It was shown in [18,25] that only density-dependent M3Y *NN* forces with the finite range exchange term allow reproducing the saturation density of cold nuclear matter. Within the framework of DFMDEF, one can calculate the interaction energy of a spherical projectile nucleus and a deformed target nucleus.

Note that in both DFMSPH and DFMDEF there is an option for treating the single-nucleon exchange (SNE) in a local approximation using the finite-range exchange part of the M3Y interaction which was proven to be a much more accurate procedure compared to the zero-range pseudopotential.

However, the density dependence of the *NN* forces absents in the DFMDEF code. To account for this density dependence in an approximate manner for the reactions with deformed target nuclei and simultaneously to reduce the computer time needed for the systematic calculations, we have developed the following algorithm.

We start from finding the barrier energies through DFMSPH (i.e. ignoring the target nucleus deformation) using the $fDDNN$ forces. The corresponding barrier energies are denoted $B_{fDDs}$ (the last "s" in the subscript stands for "spherical"). Next, we apply the $\delta NN$ density-independent forces and, varying $G_{E\delta}$, fit the Coulomb barrier energies $B_{fDDs}$. Let us denote this fitting barrier energy as $B_{\delta s}$ and the corresponding value of $G_{E\delta}$ as $G_{E\delta s}$. Then, we evaluate the $\theta$-dependent SnnP and the barrier energy $B_\delta(\theta)$, with this optimal value $G_{E\delta s}$, accounting for the deformation of the target nucleus using the code DFMDEF. These are the SnnP and barrier energy we plan to apply for calculating the cross-sections.

To verify that the resulting barrier energies $B_\delta(\theta)$ answer our purpose, we compare them with the barrier energies calculated with $fNN$ forces. For this aim, we evaluate the incident angle-dependent barrier energies $B_f(\theta)$ using the code DFMDEF accounting for the deformations of the target nucleus. Since the density dependence absents in DFMDEF, we need to find a correcting factor responsible for the contribution of the density dependence. This factor comes from the comparison of $B_{fDDs}$ and $B_{fs}$ calculated for spherical nuclei within DFMSPH using the $fNN$ forces with density dependence and without it, respectively. Thus, we obtain the corrected barrier energies $B_{fc}(\theta)$. We expect a good agreement between the simplified barrier energy $B_\delta(\theta)$ and the more rigorous $B_{fc}(\theta)$.

Note that a similar contrivance, proposed in Ref. [37], was successful but did not receive further development. In the present paper we work out that idea.

## IV. SELECTING THE REACTIONS

The nuclei and reactions for the present study have been selected according to the following considerations. First, we tried selecting the reactions in which a smaller spherical projectile nucleus collides with a significantly bigger deformed target nucleus. Second, we used only those nuclei for which in [38] the three-parameter Fermi functions (3pF-formulas) are available for the experimental charge density. Third, the reactions have been chosen to cover a wide range of the approximate Coulomb barrier

$$B_Z = \frac{Z_P Z_T}{A_P^{1/3} + A_T^{1/3}} \text{ MeV.} \quad (18)$$

Fourth, we selected the target nuclei possessing significant quadrupole and hexadecapole deformations of both signs. Finally, we prefer the reactions for which experimental data on the barrier energy and/or capture cross-sections are available. Many data are taken from the database [39].

Information on the projectile and target nuclei is comprised in Tables II and III, respectively; the reactions selected for consideration in the present work are comprised in Table IV. Even when the projectile nucleus is known to be deformed, we consider it as spherical one. This does not wrench our results due to the smaller size of the projectile nucleus.

Table II. Projectile nuclei: parameters of the experimental charge density (3pF-formula, Eq. (19)) [38].

|  | $^{16}$O | $^{19}$F | $^{20}$Ne | $^{26}$Mg | $^{27}$Al | $^{40}$Ar | $^{48}$Ti | $^{56}$Fe | $^{58}$Ni | $^{64}$Ni |
|---|---|---|---|---|---|---|---|---|---|---|
| $R_F$ | 2.608 | 2.590 | 2.805 | 3.050 | 3.070 | 3.530 | 3.843 | 4.106 | 4.309 | 4.212 |
| $a_F$ | 0.513 | 0.564 | 0.571 | 0.523 | 0.519 | 0.542 | 0.588 | 0.519 | 0.517 | 0.578 |
| $w_F$ | -0.051 | 0.000 | 0.000 | 0.000 | 0.000 | 0.000 | 0.000 | 0.000 | -0.131 | 0.000 |

Table III. Target nuclei: deformations [39,40] and parameters of the experimental charge density (3pF-formula, Eqs. (20), (21)) [38].

|  | $^{50}$Cr | $^{59}$Co | $^{65}$Cu | $^{74}$Ge | $^{154}$Sm | $^{165}$Ho | $^{181}$Ta | $^{186}$W | $^{197}$Au | $^{232}$Th | $^{238}$U |
|---|---|---|---|---|---|---|---|---|---|---|---|
| $\beta_2$ | +0.194 | +0.118 | -0.125 | -0.237 | +0.270 | +0.284 | +0.255 | +0.221 | -0.125 | +0.205 | +0.236 |
| $\beta_4$ | +0.038 | +0.005 | -0.005 | -0.036 | +0.105 | +0.020 | -0.076 | -0.095 | -0.017 | +0.103 | +0.098 |
| $R_F$ | 3.979 | 4.080 | 4.158 | 4.450 | 5.939 | 6.180 | 6.380 | 6.580 | 5.380 | 6.792 | 6.805 |
| $a_F$ | 0.520 | 0.569 | 0.632 | 0.580 | 0.522 | 0.570 | 0.640 | 0.480 | 0.535 | 0.571 | 0.605 |

The present study has a lot in common with Ref. [37]. Here comes the list of the differences: (i) in the present work, we significantly widen the range of reactions considered (20 against 7); (ii) reactions with $^{168}$Er and $^{12}$C are excluded because the parameters of the 3pF-formulas for the experimental charge densities of these nuclei are absent; (iii) the approximation for the finite-range calculations are performed here for all 20 reactions. In addition, in [37], contrary to the present work, the calculated capture cross-sections were not presented. On the other hand, the details of the calculations omitted here can be found in [36,37].

## V. NUCLEON DENSITIES

Basing on the experimental charge densities from [38], we approximate the nucleon densities for spherical nuclei by the 3pF-formula:

$$\rho_F(r) = \rho_{CF} \frac{1 - w_F r^2/R_F^2}{1 + \exp[(r - R_F)/a_F]}. \quad (19)$$

Here $a_F$ denotes the diffuseness of the density, $R_F$ approximately corresponds to the half density radius. The quantity $w_F$ seems not to have a specific name in the literature. The value of $\rho_{CF}$ is defined by the normalization condition. In [38], Eq. (19) is applied for the nuclear charge density of both spherical and deformed nuclei. In the latter case, the average over the polar angle $\zeta$ (see Fig. 1) is meant. Note that the 2pF-formula used in [38] is just the same 3pF-formula with $w_F = 0$.

The SnnP of the colliding nuclei significantly depends upon the incident angle $\theta$. Therefore, when dealing with the deformed nucleus, we use the following dependence upon $\zeta$:

$$\rho_F(r,\zeta) = \rho_{CF}\{1 + \exp[(r - R_F f(\zeta))/a_F]\}^{-1}, \quad (20)$$

$$f(\zeta) = \lambda^{-1}[1 + \beta_2 Y_{20}(\zeta) + \beta_4 Y_{40}(\zeta)]. \quad (21)$$

Here $\beta_2$ and $\beta_4$ are the parameters of the quadrupole and hexadecapole deformation, respectively, whose values for the target nuclei were taken from [40], $\lambda$ guarantees the nucleon number conservation; $Y_{20}(\zeta)$ and $Y_{40}(\zeta)$ denote the spherical functions.

The same formulas (19) and (20) are used for the charge-, neutron-, and proton densities. The half-density radii are the same for a given nucleus whereas the diffusenesses of proton $a_{Fp}$ and neutron $a_{Fn}$ densities are extracted from the experimental diffuseness $a_{Fq}$ of the charge density [1,37]:

$$a_{Fp} = a_{Fn} = \sqrt{a_{Fq}^2 - \frac{5}{7\pi^2}\left(0.76 - 0.11\frac{N}{Z}\right)}. \quad (22)$$

In the present work, the densities are considered to be frozen. This frozen density approximation (FDA) has been inspected carefully and compared with the adiabatic density approximation (ADA) in Ref. [41]. The ADA has been shown to be more appropriate at smaller center-to-center distances, in other words at larger densities overlap. We will come back to this point in Section VIA below.

## VI. RESULTS

### A. Barrier energies ignoring target deformations

All calculations in this subsection are performed through the code DFMSPH. According to the formulated algorithm, we present in this subsection the barrier energies $B_{fDDs}$ calculated using the $fDDNN$ forces but ignoring the deformations of the target nuclei. The values of $B_{fDDs}$ are shown in Table IV.

Next, we have calculated the barrier energies $B_{\delta s}$ using the $\delta NN$ forces with several values of $G_{E\delta}$ also ignoring the deformations of the target nuclei. It is convenient representing the results using the fractional difference

$$\xi_{\delta D}^{s}(B_Z) = \frac{B_{\delta s}}{B_{fDDs}} - 1. \qquad (23)$$

These differences calculated at three values of $G_{E\delta}= -940, -1040,$ and $-1140$ MeV fm³ are plotted in Fig. 2a. This figure shows that the value $G_{E\delta s} = -1040$ MeV fm³ can be taken as an optimal one for all the reactions considered. We apply this value below calling this version of calculations "modified $\delta NN$-forces".

Let us now compare the barrier energies evaluated accounting for the density dependence, $B_{fDDs}$, and without this dependence, $B_{fs}$. This comparison is presented in Fig. 2b as the fractional difference

$$\xi_{fD}^{s}(B_Z) = \frac{B_{fs}}{B_{fDDs}} - 1. \qquad (24)$$

On average, this fractional difference is about 1% (thin horizontal line), however, it reveals a tendency to increase at small values of $B_Z$ and to flatten at larger values of $B_Z$. Below we use $\xi_{fD}^{s}$ for finding a proper correcting factor for the deformed barriers. Therefore, we prefer to approximate the $\xi_{fD}^{s}(B_Z)$-dependence by the following analytical expression

$$\xi_a(B_Z) = \xi_1 \exp\left(\frac{B_0 - B_Z}{\Delta B}\right) + \xi_0. \qquad (25)$$

In Fig. 2b, $\xi_a$ calculated with $\xi_0 = 0.80$, $\xi_1 = 0.50$, $B_0 = 40$ MeV, $\Delta B = 15$ MeV is shown by the curved line.

Probably, it is worthwhile to inspect whether the FDA makes sense for the presented calculations. For this aim, we display in Fig. 3 the fractional difference

$$\mu_R(B_Z) = \frac{R_{Bfs}}{R_{FP} + R_{FT}} - 1 \qquad (26)$$

showing to what extent the spherical barrier radius calculated using the $fDDNN$ forces $R_{Bfs}$ is larger than the sum of half density radii of the projectile $R_{FP}$ and target $R_{FT}$ nuclei. This fractional difference typically exceeds 20% thus the FDA seems to be justified.

### B. Barrier energies accounting for target deformations

All calculations in this subsection are performed within the framework of the code DFMDEF. We now evaluate the barrier energies $B_\delta(\theta)$ accounting for the deformations of the target nuclei as indicated in Table III. These energies are calculated using the modified $\delta NN$ forces, i.e. with the optimal value $G_{E\delta s}$ chosen in the previous subsection. Since the value of $G_{E\delta s}$ was found by basing on the calculations with density-dependent forces, it implicitly includes this dependence. These are the calculations we intend to apply for the capture cross-section calculations. However, first we would like to test if the values of $B_\delta(\theta)$ indeed correspond to the barriers with finite-range $NN$ forces. Therefore, we calculate $B_f(\theta)$ and correct them using the correcting factor $\xi_a(B_Z)$ of Eq. (25) as follows

$$B_{fc}(\theta, B_Z) = \frac{B_f(\theta, B_Z)}{1 + \xi_a(B_Z)}. \qquad (27)$$

This procedure is extremely computer time-consuming; therefore, it is performed for several values of $\theta$ only. These barrier energies $B_{fc}$ are compared with $B_\delta(\theta)$ via the fractional difference

$$\xi_{\delta f}^{\theta}(\theta) = \frac{B_\delta(\theta)}{B_{fc}(\theta)} - 1. \qquad (28)$$

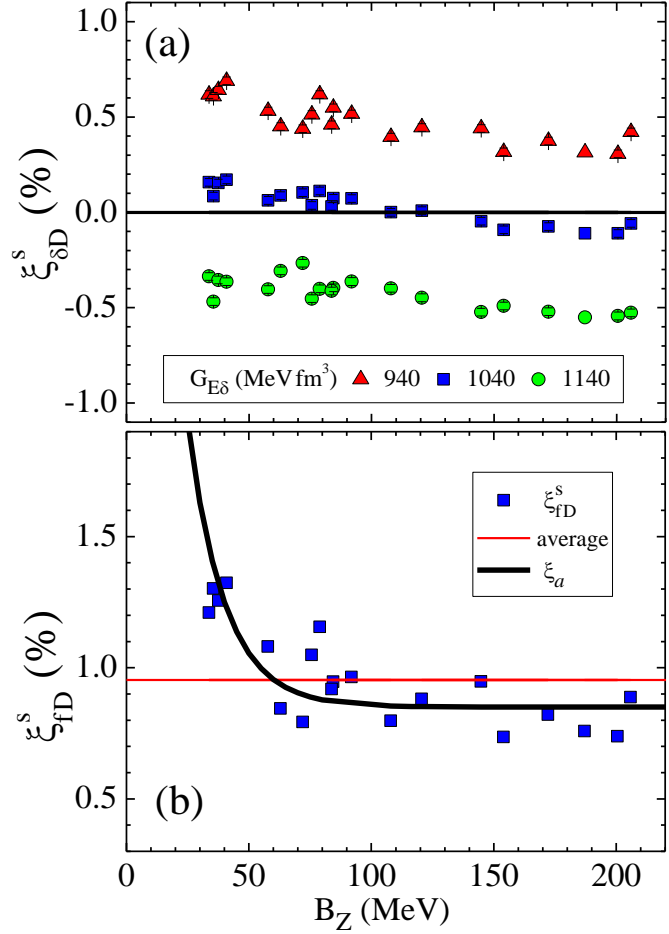

FIG. 2. Comparison of barrier energies evaluated ignoring the deformations of the target nuclei versus the approximate barrier energy $B_Z$. Panel (*a*): fractional difference $\xi^s_{\delta D}$ (see Eq. (23)) corresponding to three values of $G_{E\delta}$ indicated in the panel. Panel (b): squares correspond to fractional difference $\xi^s_{fD}$ (see Eq. (24)); thin horizontal line is the average of $\xi^s_{fD}(B_Z)$; thick black curve is the analytical approximation $\xi_a(B_Z)$ (see Eq. (25)).

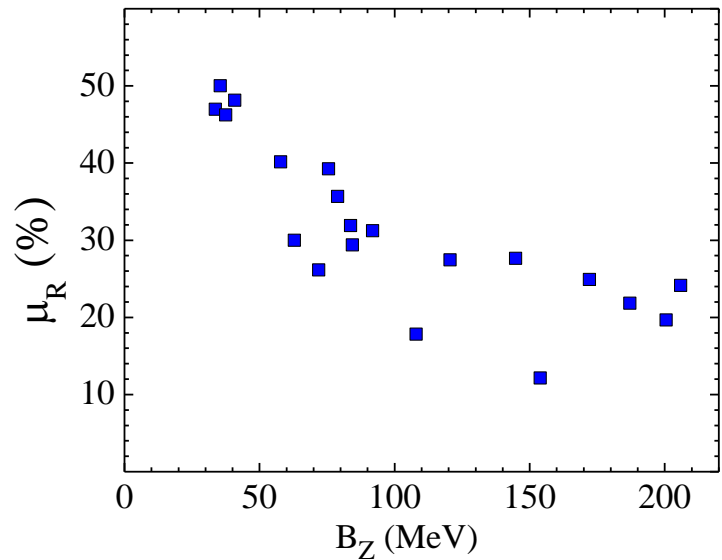

FIG. 3. Excess of the spherical barrier radius calculated using the *fDDNN* forces $R_{Bfs}$ over the sum of half density radii of the projectile $R_{FP}$ and target $R_{FT}$ nuclei versus $B_Z$ (see Eq. (26)).

Table IV. The ordering numbers of reactions, the corresponding reactions, and the barrier energies: the approximate ones $B_Z$ (Eq. (18)); $B_{\delta s}$ calculated using the $\delta NN$ forces with $G_{E\delta} = -1040$ MeV fm³; $B_{fDDs}$ evaluated using the $fDDNN$ forces; the experimental ones $B_{f\,exp}$ with the corresponding references. In the last column, the references for the experimental capture cross-sections and the number of experimental points are presented.

|     | Reaction | $B_Z$ (MeV) | $B_{\delta s}$ (MeV) | $B_{fDDs}$ (MeV) | $B_{f\,exp}$ (MeV), Reference | Number of points, capture data |
|---|---|---|---|---|---|---|
| R1  | $^{16}$O+$^{59}$Co   | 33.68  | 29.56  | 29.51  | 30.5 [42]   | 4 [42], 2 [43]        |
| R2  | $^{16}$O+$^{65}$Cu   | 35.47  | 30.66  | 30.63  | --          | 9 [44], 7 [45]        |
| R3  | $^{20}$Ne+$^{50}$Cr  | 37.51  | 32.51  | 32.46  | --          | 3 [46]                |
| R4  | $^{20}$Ne+$^{59}$Co  | 40.86  | 35.55  | 35.49  | --          | 10 [47]               |
| R5  | $^{27}$Al+$^{74}$Ge  | 57.79  | 53.14  | 53.10  | 55.2 [48]   | 18 [48]               |
| R6  | $^{16}$O+$^{154}$Sm  | 62.94  | 60.61  | 60.56  | 59.4 [49]   | 39 [49], 22 [50]      |
| R7  | $^{16}$O+$^{186}$W   | 71.95  | 69.54  | 69.47  | 59.35 [51]  | 38 [52]               |
| R8  | $^{40}$Ar+$^{74}$Ge  | 75.61  | 70.82  | 70.79  | --          | 3 [53]                |
| R9  | $^{19}$F+$^{181}$Ta  | 78.92  | 72.94  | 72.86  | --          | 10 [54], 9 [55], 8 [56] |
| R10 | $^{19}$F+$^{197}$Au  | 83.77  | 81.42  | 81.40  | 81.61 [51]  | 8 [57]                |
| R11 | $^{16}$O+$^{238}$U   | 84.43  | 81.91  | 81.85  | 85 [58]     | 4 [59], 2 [60]        |
| R12 | $^{19}$F+$^{232}$Th  | 91.91  | 89.23  | 89.17  | 89.30 [51]  | 15 [61], 6 [62]       |
| R13 | $^{26}$Mg+$^{197}$Au | 107.96 | 107.46 | 107.46 | --          | 1 [63]                |
| R14 | $^{26}$Mg+$^{238}$U  | 120.53 | 119.09 | 119.08 | 123 [58]    | 3 [63]                |
| R15 | $^{40}$Ar+$^{181}$Ta | 144.77 | 140.29 | 140.35 | --          | 6 [64]                |
| R16 | $^{40}$Ar+$^{197}$Au | 153.92 | 156.28 | 156.43 | --          | 3 [65], 3 [66], 5 [67] |
| R17 | $^{40}$Ar+$^{238}$U  | 172.19 | 173.61 | 173.74 | --          | 3 [66], 2 [68]        |
| R18 | $^{56}$Fe+$^{165}$Ho | 187.10 | 188.20 | 188.40 | --          | 4 [69]                |
| R19 | $^{58}$Ni+$^{165}$Ho | 200.52 | 202.47 | 202.69 | --          | 2 [70]                |
| R20 | $^{48}$Ti+$^{238}$U  | 205.87 | 206.90 | 207.02 | --          | 1 [60]                |

Results for all 20 reactions from Table IV are presented in Fig. 4. The deviation of the simplified barrier energies $B_\delta(\theta)$ from the more rigorous $B_{fc}(\theta)$ is within 0.5% for all values of the incident angle and $B_Z$. Thus, we believe it is justified to utilize the SnnP $U_n(R,\theta)$ with $\delta NN$ forces corresponding to the optimal value $G_{E\delta s}$ for calculating the total effective interaction energy of the colliding nuclei $U(R,\theta)$ and for evaluating the capture cross-sections.

Before doing that, we wish to discuss shortly a side result of our study, namely the scaling properties of the barrier energies $B_\delta(\theta)$. For this aim, we present in Fig. 5 one more fractional difference

$$\xi_{\delta D}^{\theta s}(\theta) = \frac{B_\delta(\theta)}{B_{fDDs}} - 1. \tag{29}$$

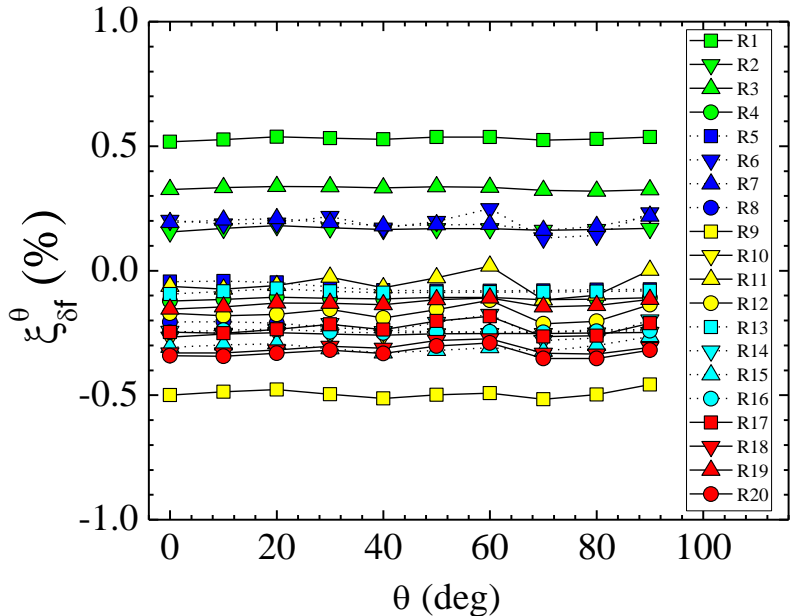

FIG. 4. Fractional difference $\xi_{\delta f}^{\theta}(\theta)$ (see Eq. (28)) versus accident angle $\theta$. For convenience, we use the ordering numbers of reactions from Table IV.

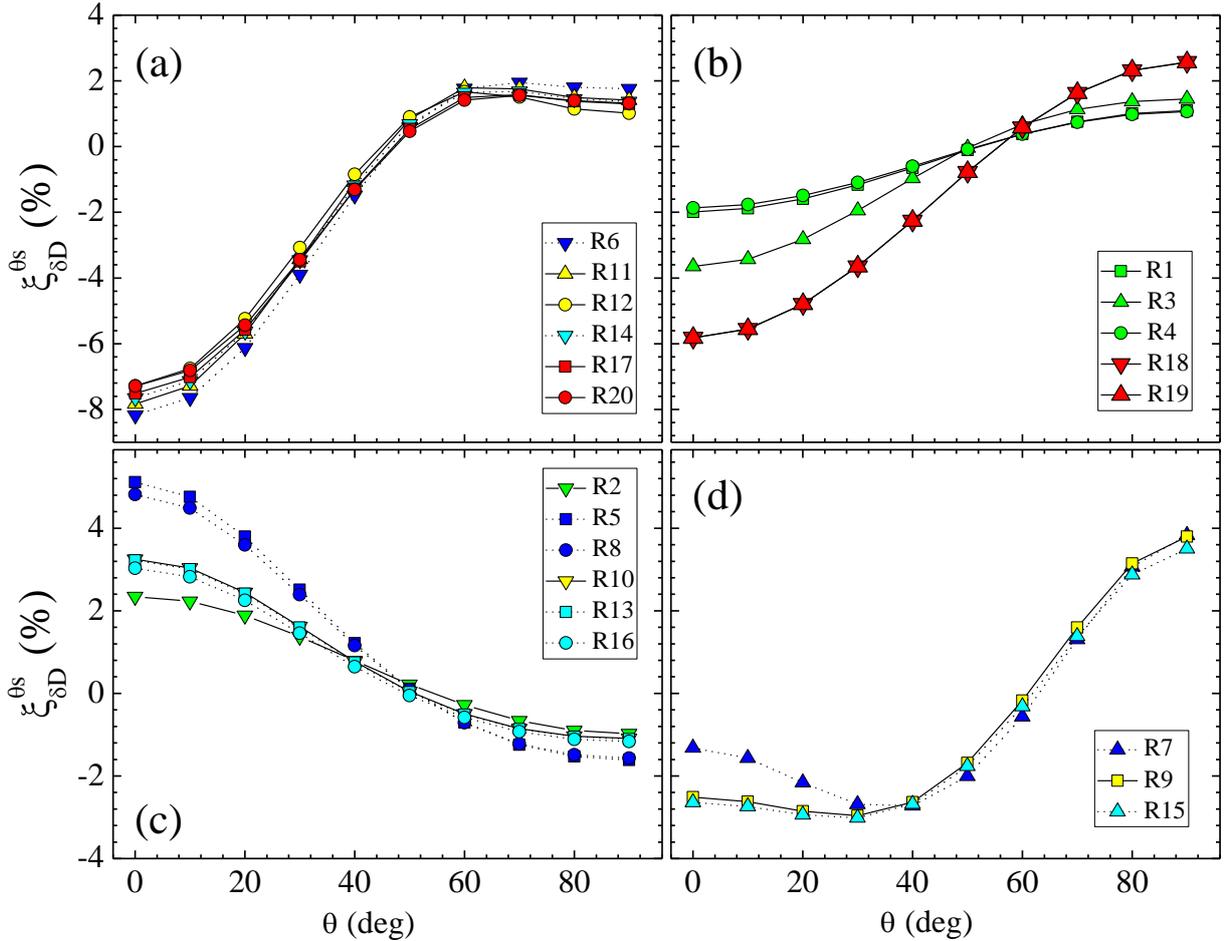

FIG. 5. Fractional difference $\xi_{\delta D}^{\theta s}(\theta)$ (see Eq. (29)) as a function of accident angle $\theta$.

In this figure, all reactions are split into four groups according to the deformations of the target nuclei: Fig. 5a comprises the cases of significant and positive $\beta_2$ and $\beta_4$; in Fig. 5b, both $\beta_2$ and $\beta_4$ are positive but $\beta_4$ is small; in Fig. 5c the cases of significant negative $\beta_2$ are collected; finally, the remaining reactions are shown in Fig. 5d. For convenience, we use the ordering numbers of reactions from Table IV. The first group is formed by the reactions with samarium, thorium, and uranium as the target nuclei whose quadrupole and hexadecapole deformations are close according to Table III. As the consequence, the curves in Fig. 5a are hardly distinguishable. Note that the absolute values of the spherical barrier energies for these reactions are quite different: from 60.6 MeV for reaction R6 up to 205.0 MeV for R20. Here the shape of $\xi_{\delta D}^{\theta s}(\theta)$ is defined by the prolate shape of the nucleus. Note that the barrier at the pole is 8% lower than the spherical barrier. This must dramatically influence the capture excitation function at the barrier region. Therefore, results of [16] where $^{238}$U and $^{246}$Cm were considered to be spherical seem to be unreliable.

Fig. 5b comprises the fractional differences $\xi_{\delta D}^{\theta s}(\theta)$ for reactions involving holmium, cobalt, and chromium target nuclei whose hexadecapole deformations are much smaller than for the first group. Accordingly, the range of $\xi_{\delta D}^{\theta s}(\theta)$ is here much narrower than in Fig. 5a. This range is especially narrow for reactions R1 and R4 involving $^{59}$Co in agreement with its comparatively small quadrupole deformation. Yet anyway the pole barrier energies are smaller than the equator ones.

The situation changes strikingly in Fig. 5c due to the opposite sign of both $\beta_2$ and $\beta_4$. Due to the oblate shape of copper, germanium, and gold nuclei, the polar barrier energies are now higher than the equatorial ones.

After the above analysis, Fig. 5d seems to be self-explanatory. A general conclusion from Fig. 5 is that the fractional difference $\xi_{\delta D}^{\theta s}(\theta)$ depends solely upon the deformations of the target nucleus. This circumstance can probably help in extracting the values of $\beta_2$ and $\beta_4$ from the experimental data.

### C. Capture cross-sections: qualitative comparison with the data

Following [37,49,71], we evaluate the capture cross-section as

$$\sigma_{th} = \frac{\pi\hbar^2}{2m_R E_{c.m.}} \sum_i \sum_J (2J+1) T_J(\theta_i) \sin(\theta_i) \Delta\theta. \tag{30}$$

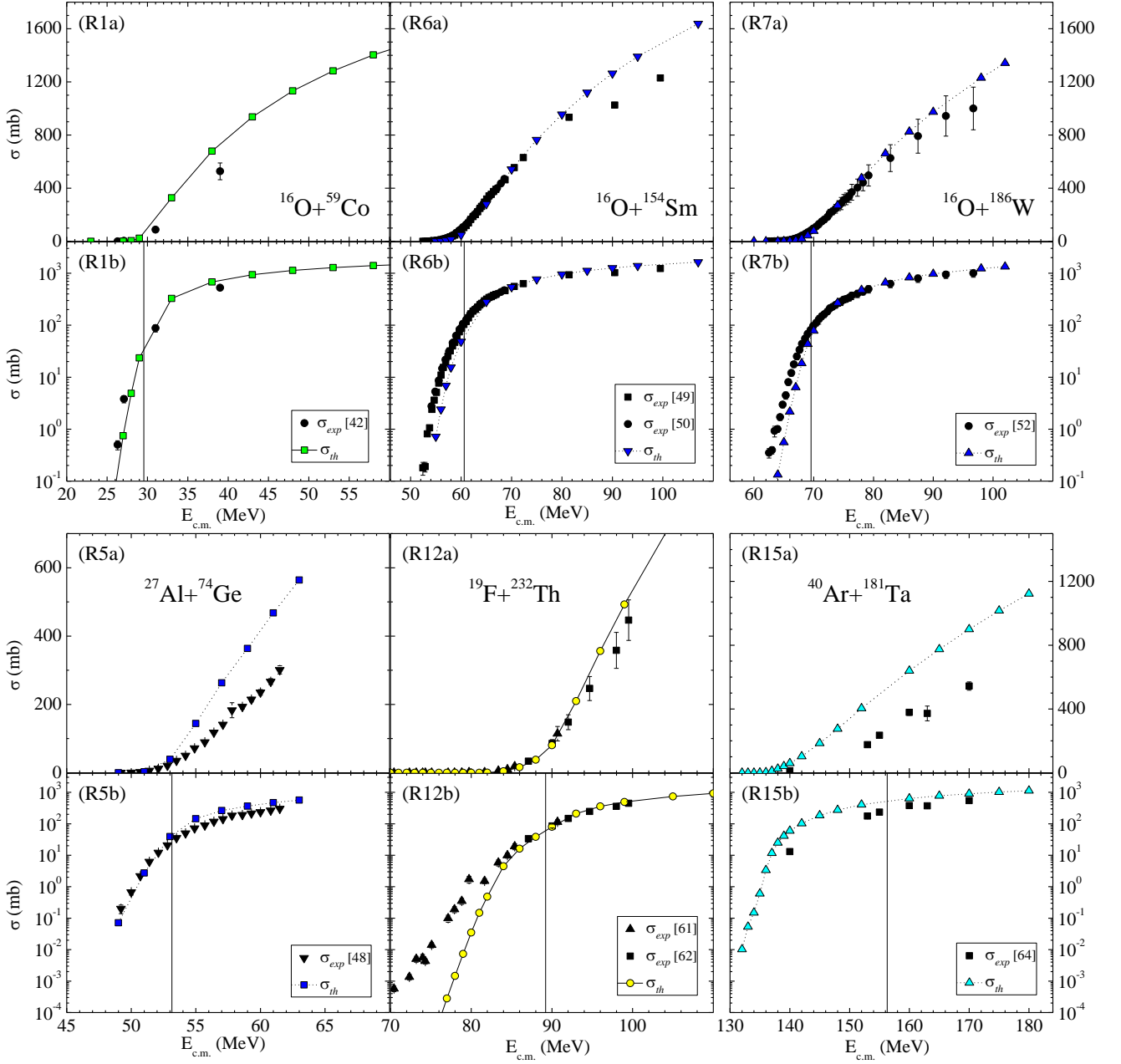

FIG. 6. Calculated and experimental cross-sections as the functions of collision energy for six reactions. The values of $B_{\delta s}$ are displayed by vertical lines. For convenience, we use the ordering numbers of reactions from Table IV.

Here, $J$ is the angular momentum in units of $\hbar$. This semi-classical approach provides results that are close to the exact quantum mechanical calculations [71,72]. The transmission coefficient is calculated using the WKB-approximation below the barrier

$$T_J(\theta) = \left\{1 + \exp\left[\frac{2S}{\hbar}\right]\right\}^{-1} \tag{31}$$

and in the parabolic barrier approximation above the barrier

$$T_J(\theta) = \{1 + \exp[2\pi(B_\delta - E_{c.m.})/(\hbar\omega_B)]\}^{-1}. \tag{32}$$

In Eq. (31), $S$ denotes the action calculated from the outer down to the inner turning points. In Eq. (32), both the barrier energy $B_\delta$ and frequency $\omega_B$ are $J$- and $\theta$-dependent. The incident angle changes with step $\Delta\theta = 2.5^o$ from $0^o$ up to $90^o$ due to the mirror symmetry of the target nuclei. The summation over $J$ is terminated when the partial cross-section becomes $10^{-5}$ of its maximum value.

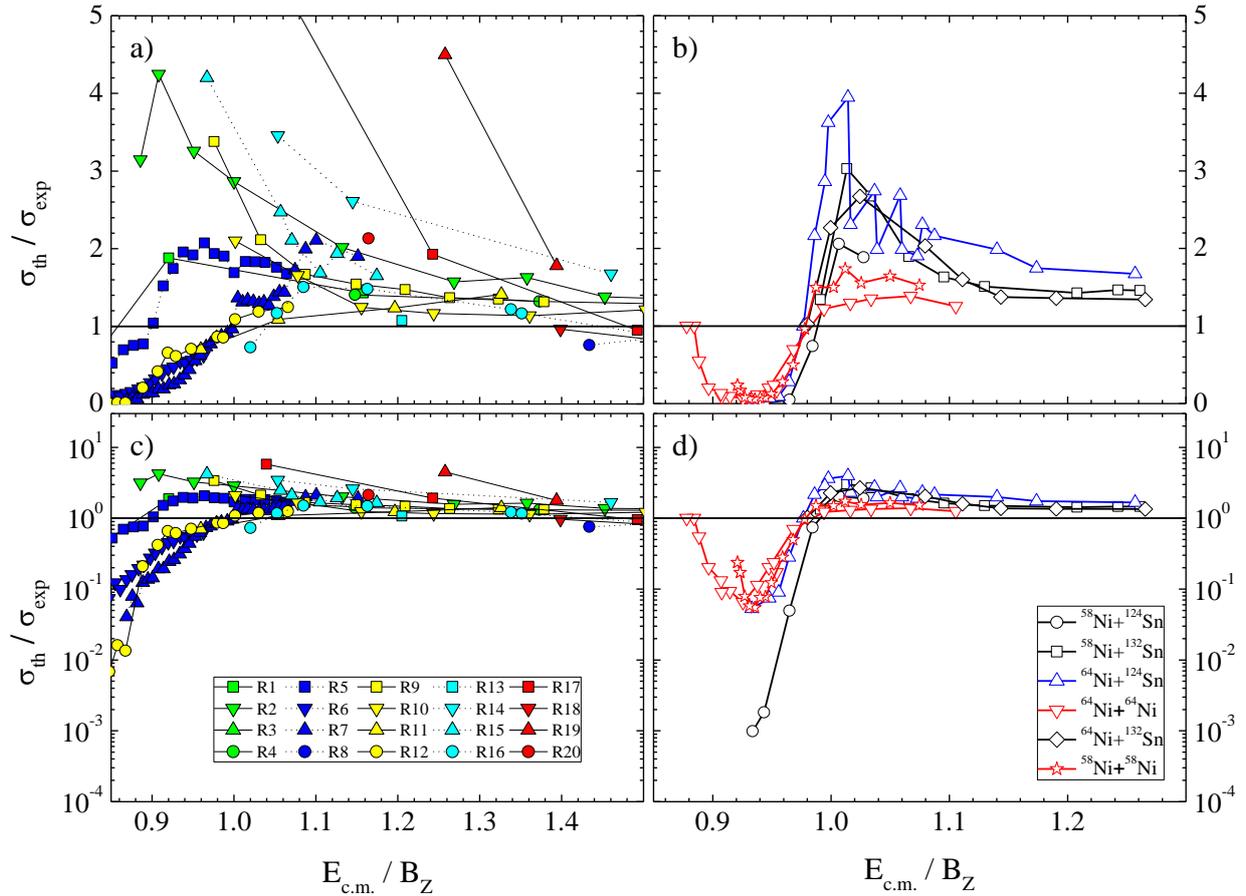

FIG. 7. Ratios of calculated cross-sections $\sigma_{th}$ and experimental $\sigma_{exp}$ as the functions of $E_{c.m.}/B_Z$. Panels (a), (c) correspond to our M3Y-calculations; panels (b), (d) are based on the RMF-calculations from [15] (the data are read from the figures of [15]). For convenience, we use the ordering numbers of reactions from Table IV.

The capture excitation functions have been calculated and compared with the data for all 20 reactions. The problem is that for many reactions only several experimental points of the excitation functions are available (see the last column of Table IV). Therefore, we have selected for the detailed presentation six reactions for which the experimental capture cross-sections $\sigma_{exp}$ obey three conditions: i) they should cover a rather wide range of $B_Z$, ii) a reasonable number of experimental points should be available, iii) for a given reaction the data should cover the barrier energy region. These data and calculated cross-sections $\sigma_{th}$ are presented in Fig. 6 (panels (a) and (b) provide the cross-sections in linear and logarithmic scales, respectively). In all cases but one, the calculated cross-sections lay below the data in the sub-barrier region and above the data at $E_{c.m.} > B_{\delta s}$. For the reactions not shown in Fig. 6, the mutual layout of $\sigma_{exp}$ and $\sigma_{th}$ do not contradict this with very few exceptions.

We interpret this observation in the following way. At $E_{c.m.} > B_{\delta s}$, the collisions become strongly dissipative and require including friction to be modelled correctly. This had been done before in many works [17,27–29,73–76]. Including dissipation within dynamical calculations can help to bring the theoretical above-barrier cross-sections in agreement with the data. Below the barrier, dissipation probably plays a minor role in comparison with the SnnP.

However, there is one more uncertainty: the deformation of the target nucleus. Here, we utilize the theoretical deformation parameters from [40] which are not always in agreement with the experimental values. For instance, for spherical $^{16}$O, Ref. [40] predicts $\beta_3 = -0.258$ and $\beta_4 = -0.122$. An alternative would be to use the experimental deformations in our calculations. Unfortunately, these data are not available for all target nuclei.

### D. Capture cross-sections: quantitative comparison with the data

To compare the calculated cross-sections $\sigma_{th}(M3Y)$ with the experimental ones more quantitatively, we collect their ratios for all analyzed reactions in Fig. 7 (in linear and logarithmic scales in panels (a) and (c), respectively) as the functions of $E_{c.m.}/B_Z$. Since the present research is motivated by the statement of [15] that the RMF $NN$ forces provide "... a better choice than the M3Y interaction for fusion reaction considered in the entire range of barrier energies in predicting the cross-sections...", in parallel, the same ratios extracted from that paper are presented in Fig. 7b and Fig. 7d. Note that in Ref. [15] only two narrow ranges of $B_Z$ were considered: 98-101 and 154-158 MeV whereas in our analysis $34 < B_Z/\text{MeV} < 206$ (see Table IV). This circumstance explains the more scattered behavior of the points in Fig. 7a and Fig. 7c. In contrast to the conclusions of the authors of [15], we do not see any superiority of the calculations performed with the RMF $NN$ forces.

# VII. CONCLUSIONS

For calculating the heavy-ion capture cross-sections, in our previous works [8,29], we successfully employed the double folding model with the M3Y *NN* forces. Recently, there appeared articles [15,16] in which the relativistic mean-field (RMF) *NN* forces were used for the same purpose. In [15], it is stated that the RMF forces provide better results in comparison with the M3Y ones. In the present work, we intended to verify this statement.

For this aim, we have performed the systematic calculations of the capture cross-sections for 20 asymmetric reactions with deformed target nuclei in a wide range of the barrier energies (34-206 MeV). To account for the density dependence of the M3Y *NN* forces, we have developed an algorithm for modifying the strength of the zero-range density-independent exchange forces. The nucleus-nucleus interaction energy has been calculated by means of the double-folding model with the nucleon densities based on the experimental charge densities. This algorithm has been validated in Sec. VIB and applied for finding the SnnP, barrier energies, and curvatures. Then, we have evaluated the transmission coefficients within the framework of the single barrier penetration model for different incident angles.

The calculated cross-sections have been compared quantitatively (i.e. point by point) with the experimental ones in Fig. 7a, c. The same has been made in Fig. 7b, d using the results of Ref. [15] (the data are read from the figures of [15]). It has turned out that the amount of agreement, i.e. the ratios $\sigma_{th}(M3Y)/\sigma_{exp}$ and $\sigma_{th}(RMF)/\sigma_{exp}$, is approximately the same. Moreover, our present analysis demonstrates that the level of agreement between $\sigma_{th}$ and $\sigma_{exp}$ might be improved by accounting for the dissipative character of the collision process above the barrier.

# APPENDIX

Here we present details of the calculations using the code DFMDEF. For the geometry of Fig.1, Eq. (1) becomes

$$U_n(R,\theta) = U_{nD}(R,\theta) + U_{nE}(R,\theta). \tag{1A}$$

The direct part is evaluated using the Fourier transform and expanding in spherical harmonics (remember, that the density dependence absents here, i.e. $F(\rho_{FA}) = 1$ in Eqs. (2), (6))

$$U_{nD}(R,\theta) = \sum_{l=0,2,\ldots}^{10} U_{nDl}(R) Y_{l0}(\theta), \tag{2A}$$

$$U_{nDl}(R) = \frac{4}{\sqrt{\pi}} \int_0^{k_{max}} dk \, k^2 j_l(kR) \tilde{v}_D(k) \tilde{A}_{PA0}(k) \tilde{A}_{TAl}(k), \tag{3A}$$

$$\tilde{A}_{P(T)Al}(k) = \int_0^{r_{max}} dr \, r^2 \rho_{P(T)Al}(r) j_l(kr), \tag{4A}$$

$$\rho_{P(T)Al}(r) = 2\pi \int_0^\pi d\zeta \sin(\zeta) \, \rho_{P(T)A}(r,\zeta) Y_{l0}(\zeta), \tag{5A}$$

$$\tilde{v}_D(k) = 4\pi \sum_{i=1}^{2} \frac{G_{Di} r_{vi}}{k^2 + r_{vi}^{-2}}. \tag{6A}$$

For evaluating the exchange part, we apply Eqs. (6)-(10). Finally, in the code the following equations are used:

$$U_{nEf}(R,\theta) = 4\pi \int_0^{s_{max}} ds \, s^2 j_0(k_{rel}s) v_{Ef}(s) G(R,s,\theta), \tag{7A}$$

$$G(R,s,\theta) = \int_0^{q_{max}} dq \, q^2 \int_0^\pi d\zeta h_P(R,s,q,\zeta,\theta) h_T(R,s,q,\zeta,\theta) \sin(\zeta), \tag{8A}$$

$$h_P(R,s,q,\zeta,\theta) = \int_0^{2\pi} d\varphi \, \rho_{PA}(p) \tilde{j}_1[k_{Peff}(p) \cdot s], \tag{9A}$$

$$h_T(R,s,q,\zeta,\theta) = \rho_{TA}(q,\zeta)\tilde{j}_1[k_{Teff}(q,\zeta)\cdot s]. \quad (10A)$$

The values of $k_{rel}$, $\tilde{j}_1$, and $k_{P(T)eff}$ are determined by Eqs. (12), (9), and (10), respectively. Here we denoted $\vec{q} = \vec{r}_T - \vec{s}/2$ and $\vec{p} = \vec{r}_P + \vec{s}/2$ (see Fig. 1). The absolute value of the last vector is equal to

$$p(R,q,\zeta,\varphi,\theta) = |\vec{q} - \vec{R}| = [q^2 + R^2 + 2Rq(\sin\theta\sin\zeta\cos\varphi - \cos\theta\cos\zeta)]^{1/2}. \quad (11A)$$

The upper limits of integration are $r_{max} = s_{max} = q_{max} = 3R_{FT}$, $k_{max} = 5$ fm$^{-1}$.

In the case of the zero-range exchange part of the *NN* interaction, Eq. (6) due to the Fourier transform reduces to the form similar to Eq. (2A) with

$$U_{nE\delta l}(R) = \frac{4}{\sqrt{\pi}}\int_0^{k_{max}} dk\, k^2 j_0(kR)\, G_{E\delta}\tilde{A}_{PA0}\tilde{A}_{TAl}. \quad (12A)$$

Note, that for the deformed case the density dependence is not applicable.